\documentclass[preprint,floats,aps,nofootinbib]{revtex4-1}

\usepackage{mathrsfs}
\usepackage{slashed}
\usepackage{color}
\usepackage{dcolumn}
\usepackage{bm}
\usepackage[caption=false]{subfig} 
\usepackage{mathrsfs,amsmath,amssymb}
\usepackage{stackrel}
\usepackage{bbding}
\usepackage{bm}
\usepackage{slashed}
\usepackage{comment}
\usepackage{soul}

\usepackage{tikz-feynman}
\usepackage{tikz}
\usetikzlibrary{decorations.pathmorphing}
\usetikzlibrary{decorations.pathmorphing} 
\tikzset{graviton/.style={decorate, decoration={snake, amplitude=.4mm, segment length=1.5mm, pre length=.5mm, post length=.5mm}, double}}

\begin{document}

\title{ Gravitational Wave Spectrum from the  Production of Dark Matter via the freeze-in Mechanism}

\author{Yonghua Wang$^{1}$}
\email{yonghuawang@mail.bnu.edu.cn}
\author{Wei Chao$^{1,2}$}
\email{chaowei@bnu.edu.cn}

\affiliation{$^1$Center of Advanced Quantum Studies, School of Physics and Astronomy, Beijing Normal University, Beijing, 100875, China \\
$^2$Key Laboratory of Multi-scale Spin Physics, Ministry of Education,
Beijing Normal University, Beijing 100875, China}

\begin{abstract}

Since the first detection of gravitational waves by ground-based interferometers, it has emerged as a novel probe for exploring physics in the early universe. The particle nature of cold dark matter (DM) and its underlying production mechanisms remain long-standing unresolved issues in the field. Notably, if DM is generated through the freeze-in mechanism in the early universe, direct laboratory detection becomes extraordinarily challenging due to its extremely weak coupling with standard model particles. In this study, we calculate the graviton bremsstrahlung process involved in the freeze-in production of dark matter, deriving the gravitational wave spectra for both the conventional freeze-in mechanism and ultraviolet freeze-in scenarios. Our analysis reveals that these spectra exhibit distinct characteristics, though they fall beyond the detection limits of currently proposed gravitational wave experiments. However, advancements in high-frequency gravitational wave detection technologies in the future may offer a means to indirectly probe the ultraviolet freeze-in mechanism.

\end{abstract}

\maketitle
\section{Introduction}
Cold dark matter (DM)~\cite{ParticleDataGroup:2024cfk},  which has been confirmed by various astrophysical observations, is one of the evidences for new physics beyond the standard model of particle physics, as the the particle nature of DM still eludes us. 
There are various DM models with mass ranging from ultra-light scales—as low as $10^{-22}$ eV, where candidates like axion-like particles\cite{Peccei:1977hh,Preskill:1982cy,Dine:1981rt,Marsh:2015xka}, dark photon\cite{Ilten:2018crw} or fuzzy dark matter\cite{Hu:2000ke, Hui:2016ltb} dominate—to super-heavy realms exceeding $10^{18}$ GeV, home to various solitons such as Q-balls\cite{Lee:1991ax,Kusenko:1997si,Kusenko:2001vu} or primordial black holes\cite{Carr:2016drx, Sasaki:2018dmp}. This extraordinary span encapsulates the theoretical diversity born from efforts to explain dark matter’s gravitational influence across cosmic scales, from the formation of galaxies and galaxy clusters to the large-scale structure of the universe.

To observe the DM in the underground laboratory, various experiments~\cite{Liu:2017drf} has been proposed. 
Weakly Interacting Massive Particles (WIMPs)\cite{jungman1996supersymmetric} are among the most thoroughly studied DM candidates. Such experiments detect the ionization signal, photon signal or phonon signal induced by the coherent elastic scattering between DM and the target nuclei or electron. 
For Axion-like particles, which was originally proposed to address the strong CP problem,  there are resonant cavity experiments~\cite{HAYSTAC:2018rwy}, Haloscope-like experiments~\cite{McAllister:2017lkb}, axion-induced birefrigence experiments. In addition, there are also many proposed novel methods that can effectively detect the ALP such as the oscillation of quantum qubit in a transmon qubit~\cite{Chen:2024aya,Chao:2024owf}, the phonon signal induced from the absorption of ALP in a superconductor~\cite{Knapen:2017ekk}, etc.     
Dark photon DM can be detected through resonant conversion into photons within a plasma environment. Many axion detectors are available for DP detection as well.
Primordial black holes (PBHs) are mainly detected through their gravitational effects. The Optical Gravitational Lensing Experiment (OGLE) is such a survey that looks for gravitational lensing events.

Although there is a wide variety of DM detectors, humanity has so far still not observed any DM signal in man-made laboratory.
Back to the production and evolution of dark matters in the early universe. Various DM has its unique production mechanism.
WIMPs are produced thermally via the so-called freeze-out mechanism \cite{zel1965magnetic, chiu1966symmetry}. Feebly interacting massive particles (FIMPs) \cite{hall2010freeze}
, which interact very weakly with the SM and have never attained thermal equilibrium, are produced  by the so-called freeze-in mechanism~\cite{Hall:2009bx}.  
Axion-like particles or dark photon can be produced via the misalignment mechanism~\cite{Nelson:2011sf}.  PBH can be produced from the density fluctuation during inflation, phase transitions in the early universe, or scalar field mediated interactions~\cite{Cai:2023uhc,Lewicki:2023ioy,Cotner:2017tir}, etc. 
From this point of view,  probing the unique signal induced by  various production mechanism is another way to identify the particle nature of dark matter.

Since the discovery of gravitational wave (GW) in the ground-based interferometers\cite{LIGOScientific:2016aoc}, GW has been regarded as a new probe for physics in the early universe due to its ability to free-stream across cosmic history. GW spectrum induced from various processes such as the decay of inflaton\cite{barman2023gravitational,Xu:2025wjq,huang2019stochastic,Barman:2023rpg,Bernal:2023wus,Xu:2024fjl,Xu:2024xmw,Bernal:2024jim,Bernal:2025lxp,Tokareva:2023mrt,Kanemura:2023pnv},  first order phase transitions~\cite{Caprini:2009yp,Chao:2017vrq,Hindmarsh:2017gnf},  have been systematically studied.  Graviton production from the SM plasma has been identified~\cite{Ringwald:2022xif,Ghiglieri:2024ghm,Ghiglieri:2015nfa,Garcia-Cely:2024ujr,ghiglieri2020gravitational,Ringwald:2020ist}. Moreover, proposals for  detecting high frequency GWs~\cite{Arvanitaki:2012cn,Goryachev:2014yra,Aggarwal:2020olq} show that high frequency GW can be a unique probe for high energy particle physics.  Considering that FIMP only feebly couple to the SM particle, it is difficult to detecting them in direct detection experiments. 
%
In this paper, we compute the GW spectrum induced by graviton bremsstrahlung during the DM  freeze-in production process. We present detailed calculations of the GW spectrum for both the infrared (IR)~\cite{Hall:2009bx} and ultraviolet (UV)~\cite{elahi2015ultraviolet} freeze-in scenarios. Our results demonstrate that both mechanisms generate high-frequency GW spectra. Notably, the GW signal arising from the UV freeze-in scenario exhibits large amplitude to be detectable by next-generation cavity-based gravitational wave experiments, offering a promising avenue for probing DM interactions in the early universe.

The remaining of the paper is organized as follows: In section II we give a brief overview on the freeze-in mechanism; Section III is devoted into the calculation of GW spectrum from the FI mechanism. The last part is concluding remarks. Details of the calculation is listed in the Appendix. 

\section{Freeze-in Mechanism}

In this section, we review the theoretical framework for DM production via the freeze-in mechanism. We will present both infrared (IR)~\cite{Hall:2009bx} and ultraviolet (UV) ~\cite{Elahi:2014fsa} freeze-in scenarios.

\subsection{IR Freeze-in Mechanism}
We consider a scenario in which a heavy scalar field \(\phi\), residing in the thermal bath of the early Universe, couples to a fermionic DM candidate \(\psi\) via a  tiny Yukawa interaction. The initial abundance of \(\psi\) is negligible at the onset of the thermal history and it never reach thermal equilibrium in the early universe due to the tiny coupling. The Lagrangian is given by
\begin{equation}
	\mathcal{L} \supset -\beta \,\phi\, \bar{\psi} \psi,
	\label{eq:IR_Lagrangian}
\end{equation}
where \(\beta\) is the dimensionless coupling constant.  

The evolution of the number density of \(\psi\) in the expanding universe is governed by the Boltzmann equation \cite{Hooper:2024avz,kolb2018early},
\begin{equation}
	\begin{aligned}
		\frac{d n_\psi}{d t} + 3 H n_\psi &= g_\phi \int d \Pi_\phi \, d\Pi_\psi \, d\Pi_{\bar \psi} \, (2\pi)^4 \delta^4(p_\phi-p_\psi-p_{\bar \psi}) \, |\mathcal{M}|^2_{\phi \to \psi \bar \psi} \, f_{\phi} \\
		&= \frac{1}{2 \pi^2} m_\phi^2 T \, K_1(m_\phi /T) \, \Gamma_{\phi \to \psi \bar \psi},
	\end{aligned}
\end{equation}
where \(H\) is the Hubble parameter, \(g_\phi=1\) being the number of internal degrees of freedom of \(\phi\), $m_\phi$ is the mass of $\phi$, \(d \Pi \equiv \frac{d^3 p}{(2\pi)^3 2E}\) denoting the Lorentz-invariant phase-space element, and \(f_\phi\) is the phase-space distribution function of \(\phi\), that is well approximated by the Boltzmann distribution \(f_\phi \simeq e^{-E_\phi /T}\). The function \(K_1\) is the first modified Bessel function of the second kind. In the limit \(m_\phi \gg m_\psi\), the decay rate \(\Gamma_{\phi \to \psi \bar{\psi}}\) takes the simple form
\begin{equation}
	\Gamma_{\phi \to \psi \bar \psi} = \frac{\beta^2 m_\phi}{8\pi}.
\end{equation}

It is convenient to express the Boltzmann equation in terms of the comoving number density \(Y_\psi \equiv n_\psi/S\), where 
\(
	S = {2 \pi^2 g_*^S T^3}/{45}
\)
being the entropy density of the universe. Making use of \(\dot{T} \approx -HT\) with 
\[
	H = \sqrt{\frac{g_*^\rho}{90}} \frac{\pi T^2}{M_P}
\]
for the radiation dominated universe, the Boltzmann equation can be rewritten as
\begin{equation}
	\frac{d Y_\psi}{d x} = \frac{90\sqrt{90}}{(2\pi)^6} \frac{M_{P}}{\sqrt{g_* ^\rho} g_* ^S} \frac{\beta^2}{m_\phi} \, x^3 K_1(x),
\end{equation}
where \(x \equiv m_\phi/T\) is a dimensionless inverse-temperature variable,  \(g_*^\rho\) and \(g_*^S\) are the effective relativistic degrees of freedom for energy density and entropy, respectively.

\begin{figure}[t]
	\centering
	\includegraphics[width=1.0\textwidth]{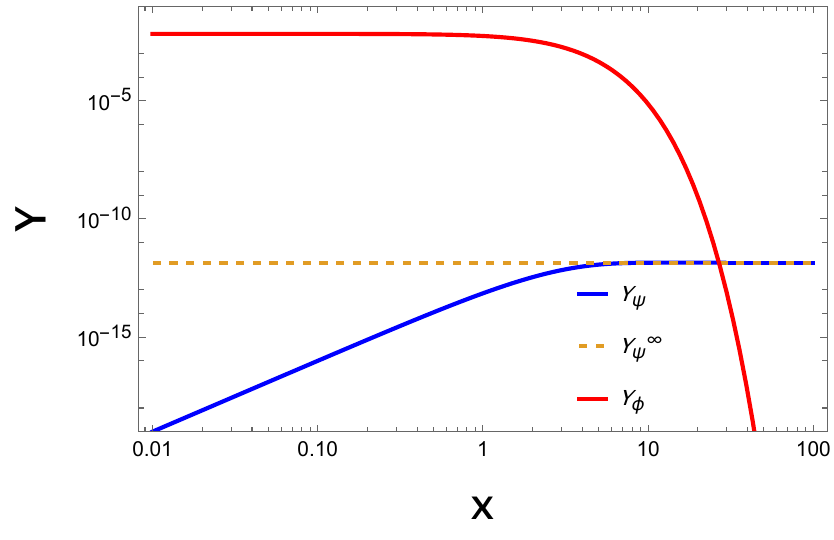}
	\caption{Evolution of the yield \(Y_\psi\) for IR freeze-in mechanism with benchmark parameters \(m_\psi = 150~\mathrm{GeV}\), \(m_\phi = 10^{14}~\mathrm{GeV}\), and \(\beta = 10^{-6}\). The red curve corresponds to the yield \(Y_\phi\) of the scalar field \(\phi\) while it is in thermal equilibrium with the plasma.}
	\label{fig:yield_evolution_IR}
\end{figure}

Integrating from \(x_i \to 0\) to \(x_f \to \infty\), the asymptotic yield of DM is
\begin{equation}
	Y_\psi^{\infty} = \frac{135\sqrt{90}}{2(2\pi)^5} \frac{M_{P}}{\sqrt{g_* ^\rho} g_* ^S} \frac{\beta^2}{m_\phi}.
\end{equation}
The present-day DM relic density \(\Omega_{\mathrm{DM}}\) is then related to \(Y_\psi^\infty\) via
\begin{equation}
	\Omega_{\mathrm{DM}} = \frac{2\,Y_\psi^{\infty} \, m_\psi \, S_0}{\rho_{cr,0}},
\end{equation}
where \(S_0\) is the entropy density today and \(\rho_{cr,0}= 8.1 \times 10^{-47} \, h^2\mathrm~{GeV}^4\) \cite{ParticleDataGroup:2024cfk} is the critical  energy density. The factor of 2 accounts for contributions from both \(\psi\) and \(\bar{\psi}\). The observed relic abundance \(\Omega_{\mathrm{DM}} h^2 \approx 0.12\)~\cite{aghanim2020planck,lahav2004cosmological, smoot2000cosmic} implies that
\begin{equation}
	\frac{\beta^2 m_\psi}{m_\phi} \simeq 1.5 \times 10^{-24} \; ,
	\label{eq:IR_constraint}
\end{equation}
and the corresponding relic abundance can be written as
\begin{equation}
	\Omega_{\mathrm{DM}} h^2 \simeq 0.12 \times \left(\frac{m_\psi}{150~\mathrm{GeV}}\right) \left(\frac{\beta}{10^{-6}}\right)^2 \left(\frac{10^{14}~\mathrm{GeV}}{m_\phi}\right).
\end{equation}

As an illustration, we set \(m_\psi = 150~\mathrm{GeV}\), \(m_\phi = 10^{14}~\mathrm{GeV}\), and \(\beta = 10^{-6}\).  Figure~\ref{fig:yield_evolution_IR} shows the evolution of the yield \(Y_\psi\) for the IR freeze-in  mechanism with  above benchmark parameters. The red curve illustrates the evolution of the yield \(Y_\phi\) of the scalar field \(\phi\) while it remains in thermal equilibrium with the plasma. The blue solid line denotes the evolution of $Y_\psi$ and the dashed horizontal line represents the required $Y_\psi$ to address the observed DM relic density.

\subsection{UV freeze-in mechanism}
We proceed to review the ultraviolet (UV) freeze-in mechanism by taking the DM $\chi$ as a complex scalar field. This field interacts only feebly with the thermal bath via high-dimensional effective operators, ensuring that $\chi$ never thermalizes with the primordial plasma. Specifically, we consider a dimension-five effective interaction between $\chi$ and a fermion species $f$ in the thermal bath:
\begin{equation}
	\mathcal{L} \supset -\frac{1}{\Lambda} \, \chi \, \chi^{\dagger} \, \bar{f} f , \label{masterxxx}
\end{equation}
where \(\Lambda\) is the cut-off scale, related to a high-energy theory beyond the SM.  

In this setup, the dominant production channel for \(\chi\) is the \(2 \to 2\) scattering process \(f \, \bar{f} \to \chi \, \chi^{\dagger}\) mediated by the  contact operator in Eq. (\ref{masterxxx}). The number density of \(\chi\) is governed by the following Boltzmann equation
\begin{equation}
	\frac{d n_\chi}{dt} + 3 H n_\chi 
	= g_f g_{\bar{f}}\int d\Pi_f \, d\Pi_{\bar{f}} \, d\Pi_\chi \, d\Pi_{\chi^\dagger} \, (2\pi)^4 \delta^4(p_f + p_{\bar{f}} - p_\chi - p_{\chi^\dagger}) \, |\mathcal{M}|^2_{f \bar{f} \to \chi \chi^\dagger} \, {\cal F}_f \, {\cal F}_{\bar{f}},
\end{equation}
where \(d \Pi_i \) is the Lorentz-invariant phase-space element, \({\cal F}_f\) and \({\cal F}_{\bar{f}}\) are the phase-space distribution functions of the fermions in the thermal bath, and \(|\mathcal{M}|^2\) denotes the squared scattering amplitude, summed over final-state and averaged over initial-state internal degrees of freedom.

In the relativistic limit, the squared amplitude averaged over initial spins and summed over final states takes the following form
\begin{equation}
	|\mathcal{M}|^2_{f \bar{f} \to \chi \chi^\dagger} \simeq \frac{s}{2\Lambda^2},
\end{equation}
where \(s\) is the Mandelstam variable corresponding to the square of the center-of-mass energy.  
Assuming Maxwell–Boltzmann statistics for the initial-state fermions, the collision term can be simplified to \cite{hall2010freeze}
\begin{equation}
	\begin{aligned}
		\frac{d n_\chi}{dt} + 3 H n_\chi \simeq \frac{T}{1024\pi^5} \int_{0}^{\infty} ds \, s^{3/2} K_1\left(\frac{\sqrt{s}}{T}\right) \frac{1}{\Lambda^2} 
		=\frac{T^6}{ (2\pi)^5 \Lambda^2}
	\end{aligned}
\end{equation}
where \(K_1\) is the first modified Bessel function of the second kind.

\begin{figure}[t]
	\centering
	\includegraphics[width=1.0\textwidth]{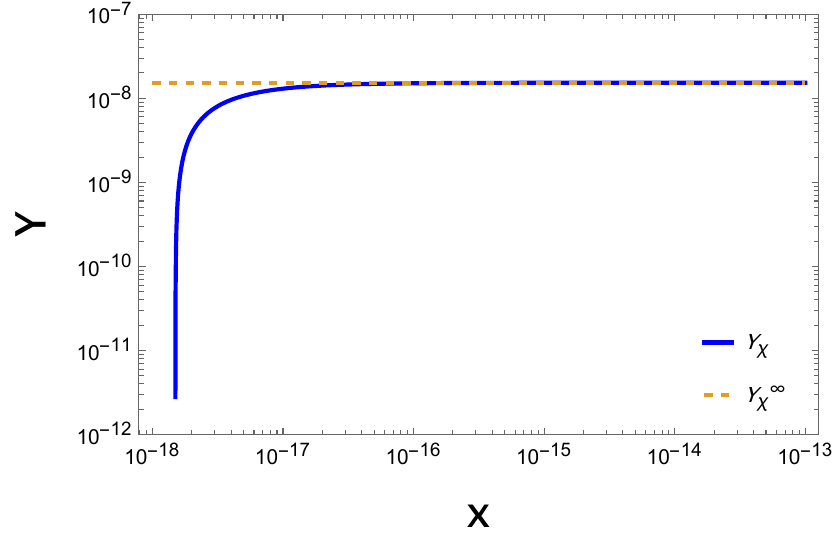}
	\caption{Evolution of the yield \(Y_\chi\) for UV freeze-in DM production with benchmark parameters \(m_\chi = 15~\mathrm{MeV}\), \(\Lambda = 10^{18}~\mathrm{GeV}\), and \(T_R = 10^{16}~\mathrm{GeV}\), where \(x\equiv m_\chi/T\).}
	\label{fig:yield_evolution_UV}
\end{figure}

Introducing the yield variable \(Y_\chi \equiv n_\chi / S\) and using \(\dot{T} \simeq -HT\), the Boltzmann equation can be recasted as
\begin{equation}
	\frac{d Y_\chi}{d T} = -\frac{45 \sqrt{90}}{2\pi^3 (2\pi)^5}\frac{1}{g_* ^S \sqrt{g_* ^\rho}}\frac{M_P}{\Lambda^2}
\end{equation}
Integrating this equation from \(T_i = T_
R\) to \(T_f = 0\), we obtain the asymptotic yield of DM,
\begin{equation}
	Y_\chi^\infty \simeq \frac{180 \sqrt{90}}{(2\pi)^8}\frac{1}{g_* ^S \sqrt{g_* ^\rho}}\frac{M_P T_R}{\Lambda^2}
\end{equation}
where \(T_{R}\) is the reheating temperature after inflation.
%
%
The observed relic abundance \(\Omega_{\mathrm{DM}} h^2 \approx 0.12\) imposes the  following constraint 
\begin{equation}
	\frac{m_\chi T_R}{\Lambda^2} \simeq 1.5 \times 10^{-22} \; ,
	\label{eq:UV_constraint}
\end{equation}
and the DM relic abundance can be written as
\begin{equation}
	\Omega_{\mathrm{DM}} h^2 \simeq 0.12 \times \left(\frac{m_\chi}{15~\mathrm{MeV}}\right) \left(\frac{T_R}{10^{16}~\mathrm{GeV}}\right) \left(\frac{10^{18}~\mathrm{GeV}}{\Lambda}\right)^2.
\end{equation}

As an illustration, we set \(m_\chi = 15~\mathrm{MeV}\), \(\Lambda = 10^{18}~\mathrm{GeV}\), and \(T_R = 10^{16}~\mathrm{GeV}\). 
Figure \ref{fig:yield_evolution_UV} shows the evolution of the yield \(Y_\chi\) for the UV freeze-in mechanism with physical meanings of each curves addressed in the caption.  Obviously, the DM is mainly produced at near the reheating temperature.

\section{GW Spectrum}
GWs can be emitted through graviton bremsstrahlung processes \cite{Weinberg:1965nx}, and are therefore expected to accompany the production of DM in the early Universe \cite{Konar:2025iuk}. In this section, we investigate such GW signals in the context of the freeze-in mechanism, where DM is produced via extremely feeble interactions with the thermal bath. Depending on the nature of the interaction, the freeze-in process can be either infrared (IR) dominated—driven by low-temperature dynamics through renormalizable operators—or ultraviolet (UV) dominated, involving higher-dimensional operators effective at high temperatures. We perform a detailed computation of the GW spectra generated in both scenarios, highlighting the distinct spectral features and their dependence on the underlying particle physics parameters. Our results provide insight into the potential observability of these signals through future high-frequency gravitational wave detectors and cosmological probes.

\subsection{GWs from IR Freeze-in DM production}

In this section, we calculate the GW spectrum arising from graviton bremsstrahlung during the production of DM via the IR freeze-in mechanism. We take the DM as $\psi$ and take the parent particle as $\phi$, with interaction given in the Eq.~\ref{eq:IR_Lagrangian}. Relevant Feynman diagrams for the graviton bremsstrahlung are listed in Fig~\ref{fig:IR_freeze_in_diagram} with Feynman rules given in the Appendix A.

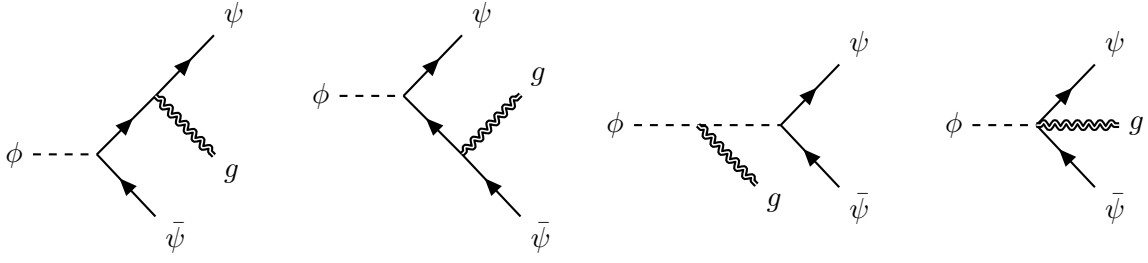
\begin{figure}[t]
	\centering
\begin{minipage}[c]{.24\textwidth}
		\begin{tikzpicture}
			\begin{feynman}
				\vertex (a) {\(\phi\)};
				\vertex [right=1.1cm of a] (b) ;
				\vertex [above right=1.1cm of b] (c);
				\vertex [above right=1.1cm of c] (d) {\(\psi\)};
				\vertex [below right=1.1cm of c] (e) {\(g \)};
				\vertex [below right=1.1cm of b] (f) {\(\bar \psi \)};
				\diagram* {
					(a) -- [scalar,thick] (b), 
					(b) -- [fermion,thick] (c),
					(c) -- [fermion,thick] (d),
					(c) -- [graviton,thick] (e),
					(f) -- [fermion,thick] (b)
				};
				~~~~~~\end{feynman}
		\end{tikzpicture}
\end{minipage}	
	\begin{minipage}[c]{.24\textwidth}
		\begin{tikzpicture}
			\begin{feynman}
				\vertex (a) {\(\phi\)};
				\vertex [right=1.1cm of a] (b) ;
				\vertex [above right=1.1cm of b] (c) {\(\psi\)};
				\vertex [below right=1.1cm of b] (d) ;
				\vertex [below right=1.1cm of d] (f) {\(\bar \psi \)};
				\vertex [above right=1.1cm of d] (e) {\(g \)};
				\diagram* {
					(a) -- [scalar,thick] (b), 
					(b) -- [fermion,thick] (c),
					(d) -- [fermion,thick] (b),
					(d) -- [graviton,thick] (e),
					(f) -- [fermion,thick] (d)
				};
				~~~~~~\end{feynman}
		\end{tikzpicture}
	\end{minipage}
\begin{minipage}[c]{.24\textwidth}
	\begin{tikzpicture}
		\begin{feynman}
			\vertex (a) {\(\phi\)};
			\vertex [right=1.1cm of a] (b) ;
			\vertex [right=1.1cm of b] (d) ;
			\vertex [below right=1.1cm of b] (c) {\(g\)};
			\vertex [above right=1.1cm of d] (e) {\(\psi\)};
			\vertex [below right=1.1cm of d] (f) {\(\bar\psi\)};
			\diagram* {
				(a) -- [scalar,thick] (b), 
				(b) -- [scalar,thick] (d),
				(b) -- [graviton,thick] (c),
				(d) -- [fermion,thick] (e),
				(f) -- [fermion,thick] (d)				
                };
			~~~~~~\end{feynman}
	\end{tikzpicture}
\end{minipage}
	\begin{minipage}[c]{.24\textwidth}
		\begin{tikzpicture}
			\begin{feynman}
				\vertex (a) {\(\phi\)};
				\vertex [right=1.1cm of a] (b) ;
				\vertex [above right=1.1cm of b] (c){\(\psi \)};
				\vertex [right=1.1cm of b] (d) {\(g \)};
				\vertex [below right=1.1cm of b] (e) {\(\bar\psi \)};
				\diagram* {
					(a) -- [scalar,thick] (b), 
					(b) -- [fermion,thick] (c),
					(e) -- [fermion,thick] (b),
					(b) -- [graviton,thick] (d)
				};
				~~~~~~\end{feynman}
		\end{tikzpicture}
	\end{minipage}	
	\caption{Feynman diagrams for graviton bremsstrahlung in the DM production process via the freeze-in mechanism. }
	\label{fig:IR_freeze_in_diagram}
\end{figure}

The squared amplitude for this process is given by:
\begin{equation}
    |\mathcal{M}|^2  = \frac{\beta^{2} m_{\phi}^{2}}{2M_{P}^{2}}\left(1-\frac{2 \omega}{m_{\phi}}\right)\left[2-\frac{2 m_{\phi}}{\omega}+\left(\frac{m_{\phi}}{\omega}\right)^{2}\right]
\end{equation}
where $\omega$ is the energy of the graviton. Here, we have assumed that the mass of the scalar field \(\phi\) is much larger than that of the fermion \(\psi\), i.e., \(m_{\phi} \gg m_{\psi}\). Thus, the mass of the fermion can be neglected in the calculation. Details of the calculation are provided in the appendix \ref{sec:squared amplitude}.

The evolution of the GW energy density is governed by the  following Boltzmann equation:
\begin{equation}
    \frac{d \rho_{GW}}{d t} + 4 H \rho_{GW} = C_{\phi \rightarrow \psi \bar{\psi} g} 
\end{equation}
where \( C_{\phi \rightarrow \psi \bar{\psi} g} \) is the collision term for the process, 
\begin{equation}
    C_{\phi \rightarrow \psi \bar{\psi} g} = g_\phi \int d\Pi_\phi d\Pi_\psi d\Pi_{\bar{\psi}} d\Pi_g (2\pi)^4 \delta^4(p_\phi - p_\psi - p_{\bar{\psi}} - p_g) |\mathcal{M}|^2 f_\phi \omega
\end{equation}
where  \( f_\phi \) is the distribution function of the particle \( \phi \),  assumed to be the Boltzmann distribution, i.e, \( f_\phi =\mathrm{e}^{-E_\phi/T} \). 
We have neglected the collision term for the back-reaction, as the number density of the fermion \(\psi\) and \( \bar \psi \) are negligible in the freeze-in process. 
Integrating over all the phase space except for the energy of the graviton, one obtains the differential collision term as
\begin{equation}
    \frac{d}{d \omega}C_{\phi \rightarrow \psi \bar{\psi} g} =\frac{m_\phi T}{ 4(2\pi)^5}  K_1 (m_\phi /T )  \omega^2  |\mathcal{M}|^2
\end{equation}
where \( K_1 \) is the first modified Bessel function of the second kind, arising from the integration over the momentum of the scalar field \(\phi\). We present the details of the integration in the appendix \ref{sec:collision_terms}. Defining the yield of the gravitons as \( Y_{GW} \equiv \rho_{GW}/S^{4/3}\) and  the variable \(x\equiv m_\phi/T\), one can rewrite the Boltzmann equation as:
\begin{equation}
    \begin{aligned}
		\frac{d}{d x}\left( \frac{d Y_{GW}}{d\omega} \right) &= \frac{T^2}{m_\phi}\frac{1}{H T S^{4/3}} \frac{d C_{\phi \rightarrow \psi \bar{\psi} g}}{d \omega}
		= \left( \frac{45}{2\pi^2 g_*^S(T)} \right)^{4/3} \frac{\beta^2 \sqrt{90}}{4(2\pi)^6\sqrt{g_*^\rho(T)}} \frac{\omega_D ^2}{M_P T_D^2}\\ &~\times K_1 \left(x \right) x^2 \left(1-\frac{2 \omega_D }{ T_D}\frac{1}{x}\right)\left[2-\frac{2 T_D }{\omega_D}x+\left(\frac{T_D}{\omega_D}x\right)^{2}\right]
	\end{aligned}
\end{equation}
where \( M_P = \sqrt{1/(8\pi G)} \) is the reduced Planck mass, \( T_D \) is the temperature at which graviton production ceases, corresponding to the time when all \(\phi\) particles have decayed, typically at \( m_\phi / T_D \sim 25 \), i.e., \( T_D \approx 0.04\, m_\phi \), and \( \omega_D \) denotes the energy of gravitons at the temperature \( T_D \). We took into account the energy redshift of the graviton in the above equation, which is given by \cite{Ringwald:2020ist}
\begin{equation}
    \omega(T) = \omega_D \frac{a(T_D)}{a(T)} = \left(\frac{g_*^S(T)}{g_*^S(T_D)}\right)^{1/3} \frac{T}{T_D} \omega_D
\end{equation}
where \(g_*^S(T_D) = 106.75\) is the effective number of relativistic degrees of freedom at the temperature \(T_D\). Before the particle \(\phi\) becomes non-relativistic, \(g_*^S(T) = 107.75\). It is therefore a good approximation to take \(g_*^S(T)/g_*^S(T_D) \simeq 1\). The graviton production begins at the reheating temperature \(T_R\) at which \( x_i=m_\phi /T_R \simeq 0 \), and ends at the temperature $T_D$ at which \( x_f=m_\phi /T_D \simeq 25\). Since \( Y_{GW}(x=T_D) \simeq Y_{GW}(x=\infty) \), we can integrate the above equation from \( x_i=0\) to \( x_f = \infty\) to obtain the yield of the GW per energy interval at the temperature of \(T_D\):
\begin{equation}
   \frac{dY_{GW}^{\infty}}{d\omega_D} = \left( \frac{45}{2\pi^2 g_*^S(T_D)} \right)^{4/3} \frac{\beta^2 \sqrt{90}}{2(2\pi)^6\sqrt{g_*^\rho(T_D)}M_P} \left( 8-3\pi \frac{\omega_D}{T_D}+6\frac{\omega_D^2}{T_D^2}-\pi \frac{\omega_D^3}{T_D^3} \right)
\end{equation}
After \(T_D\), the yield of the GW remains constant, and the energy of gravitons redshifts as the universe expands. Using \( \omega = 2\pi f \) with \(f\) being the frequency of the GW, we obtain the present-day GW spectrum as:
\begin{equation}
    \begin{aligned}
        \Omega_{GW}(f) \equiv& \frac{1}{\rho_{cr,0}}\frac{d \rho_{GW}(T_0)}{d \ln \omega_0}= \frac{1}{\rho_{cr,0}}  \frac{d Y_{GW}^\infty}{d\ln \omega_D} S^{4/3} (T_0) \\
        =& \frac{T_0^4}{\rho_{cr,0}}  \frac{g_*^S (T_0)}{g_*^S (T_D)}  \frac{ \sqrt{90}\beta^2 T_D}{2(2\pi)^5\sqrt{g_*^\rho(T_D)}M_P} \frac{f}{T_0} \times \\
		& \left[ 8-6\pi^2 \frac{f}{T_0}\left(\frac{g_*^S(T_D)}{g_*^S(T_0)}\right)^{1/3}+24\pi^2 \frac{f^2}{T_0 ^2} \left(\frac{g_*^S(T_D)}{g_*^S(T_0)}\right)^{2/3}-8\pi^4 \frac{f^3}{T_0^3}\frac{g_*^S(T_D)}{g_*^S(T_0)} \right]
		\label{eq:GW_spectrum_IR}
    \end{aligned}
\end{equation}
where \(\rho_{cr,0} = 8.1 \times 10^{-47} \, h^2\mathrm{GeV}^4\) \cite{ParticleDataGroup:2024cfk} being the critical density of the universe, and \(T_0 = 2.7 \, \mathrm{K} = 2.3 \times 10^{-13}\,\mathrm{GeV} \) being the temperature of the universe today. The effective number of relativistic degrees of freedom at the present time is \(g_*^S(T_0) = 3.91\).
\begin{figure}[t]
    \centering
		\includegraphics[width=1.0\textwidth]{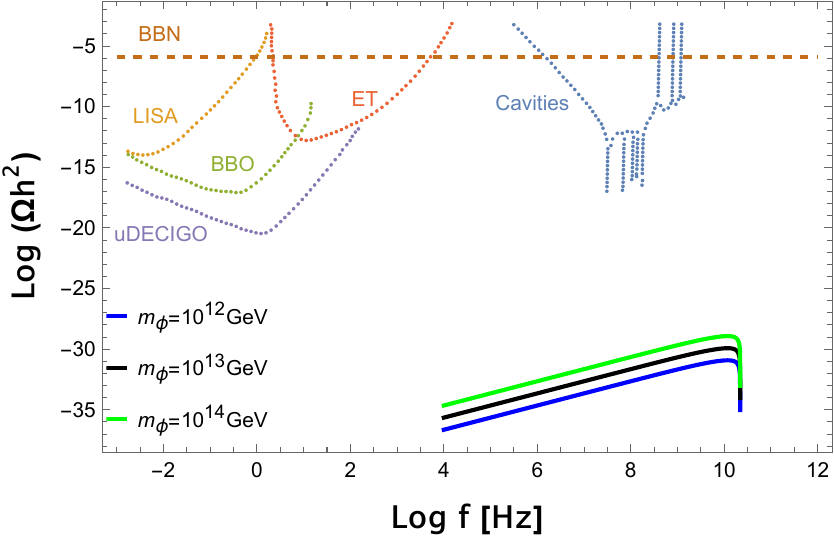}
   \caption{
The present-day GW spectrum from graviton bremsstrahlung during DM  production via freeze-in mechanism. The spectra are shown for different values of the mass of the particle \(\phi\) with fixed \(\beta = 10^{-6}\): blue (\(m_\phi = 10^{12}\,\mathrm{GeV}\)), black (\(m_\phi = 10^{13}\,\mathrm{GeV}\)), and green (\(m_\phi = 10^{14}\,\mathrm{GeV}\)). The corresponding DM candidates mass are \(m_\psi = 1.5 GeV\) (blue), \(m_\psi = 15 GeV\) (black), and \(m_\psi = 150 GeV\) (green). 
   }
    \label{fig:GW_spectrum_IR}
\end{figure}

The energy of gravitons ranges from \(0\) to \(m_\phi/2\), so the maximum frequency of the GW is
\begin{equation}
	\begin{aligned}
		f_{max} &= \frac{1}{2\pi} \omega_D^{max}\frac{a(T_D)}{a(T_0)} = \frac{1}{2\pi} \frac{m_\phi}{2} \left(\frac{g_*^S(T_0)}{g_*^S(T_D)} \right)^{1/3} \frac{T_0}{T_D} \simeq 2.3\times 10^{11} \mathrm{Hz} 
	\end{aligned}
\end{equation}
where we have used \(T_D\simeq 0.04 m_\phi\). As shown, the maximum frequency is independent of the mass of \(\phi\). According to Eq.~\ref{eq:GW_spectrum_IR}, the peak frequency of the GW spectrum is \(f_{\text{peak}} = 1.2 \times 10^{10}\,\mathrm{Hz}\), and the maximum value of the spectrum \(\Omega_{GW}(f_{peak}) \propto T_D\beta^2 \propto m_\phi \beta^2 \). The GW spectrum evaluated at the peak frequency is
\begin{equation}
	\Omega_{GW} h^2 (f_{peak}) = 1.2 \times 10^{-29} \left( \frac{\beta}{10^{-6}} \right)^2 \frac{m_\phi}{10^{14}GeV}
\end{equation}

To reproduce the observed DM relic density, \(\Omega_{DM} h^2 = 0.12\), the mass of the particle \(\phi\), the coupling constant \(\beta\), and the mass of the particle \(\psi\) (assumed to be the DM candidate) must satisfy the following relation: \( \beta^2 m_\psi / m_\phi \approx 1.5 \times 10^{-24}\). We fix the value of \(\beta = 10^{-6}\) and vary the mass of the particle \(\phi\) to obtain the GW spectrum, which is shown in the Figure \ref{fig:GW_spectrum_IR}.  And we compare the GW spectrum with the sensitivity of various experiments, including LISA \cite{amaro2017laser}, the Einstein Telescope (ET) \cite{punturo2010einstein, hild2011sensitivity, maggiore2020science, sathyaprakash2012scientific}, the Big Bang Observer (BBO) \cite{corbin2006detecting, harry2006laser, crowder2005beyond}, ultimate DECIGO (uDECIGO) \cite{seto2001possibility, kudoh2006detecting}, and resonant cavities \cite{berlin2022detecting,berlin2023electromagnetic,herman2023electromagnetic}. The constraint from the BBN,
 \( \Omega_{GW}h^2 < 1.3\times10^{-6}\)\cite{yeh2022probing}, is depicted by the dashed line.  Apparently, the GW from this process is too weak to be probed in the projected GW detection experiments.

\subsection{GW from UV Freeze-in mechanism}
\begin{figure}[t]
	\centering
	\begin{minipage}[c]{.24\textwidth}
		\begin{tikzpicture}
			\begin{feynman}
				\vertex (o) ;
				\vertex [above left= 1.1cm of o] (a) ;
				\vertex [below left= 1.1cm of o] (b) {\(\bar f\)};
				\vertex [above right= 1.1cm of o] (c) {\(\chi\)};
				\vertex [below right= 1.1cm of o] (d) {\(\chi^\dagger\)};
				\vertex [above left= 1.1cm of a] (e) {\(f\)};
				\vertex [below left= 1.1cm of a] (g) {\(g\)}; 
				\diagram* {
					(a) -- [fermion,thick] (o),
					(o) -- [fermion,thick] (b),
					(o) -- [scalar,thick] (c),
					(o) -- [scalar,thick] (d),
					(e) -- [fermion,thick] (a),
					(g) -- [graviton,thick] (a)
				};
				~~~~~~\end{feynman}
		\end{tikzpicture}
	\end{minipage}	
	\hspace{0.5cm}
	\begin{minipage}[c]{.24\textwidth}
		\begin{tikzpicture}
			\begin{feynman}
				\vertex (o) ;
				\vertex [above left= 1.1cm of o] (a) {\(f\)};
				\vertex [below left= 1.1cm of o] (b) ;
				\vertex [above right= 1.1cm of o] (c) {\(\chi\)};
				\vertex [below right= 1.1cm of o] (d) {\(\chi^\dagger\)};
				\vertex [below left= 1.1cm of b] (e) {\(\bar f\)};
				\vertex [above left= 1.1cm of b] (g) {\(g\)}; 
				\diagram* {
					(a) -- [fermion,thick] (o),
					(o) -- [fermion,thick] (b),
					(o) -- [scalar,thick] (c),
					(o) -- [scalar,thick] (d),
					(b) -- [fermion,thick] (e),
					(g) -- [graviton,thick] (b)
				};
				~~~~~~\end{feynman}
		\end{tikzpicture}
	\end{minipage}
	\hspace{0.5cm}
	\begin{minipage}[c]{.24\textwidth}
		\begin{tikzpicture}
			\begin{feynman}
				\vertex (o) ;
				\vertex [above left= 1.1cm of o] (a) {\(f\)};
				\vertex [below left= 1.1cm of o] (b) {\(\bar f\)};
				\vertex [above right= 1.1cm of o] (c) ;
				\vertex [below right= 1.1cm of o] (d) {\(\chi^\dagger\)};
				\vertex [above right= 1.1cm of c] (e) {\(\chi\)};
				\vertex [below right= 1.1cm of c] (g) {\(g\)}; 
				\diagram* {
					(a) -- [fermion,thick] (o),
					(o) -- [fermion,thick] (b),
					(o) -- [scalar,thick] (c),
					(o) -- [scalar,thick] (d),
					(e) -- [scalar,thick] (c),
					(g) -- [graviton,thick] (c)
				};
				~~~~~~\end{feynman}
		\end{tikzpicture}
	\end{minipage}
	\hspace{0.5cm}
	\begin{minipage}[c]{.24\textwidth}
		\begin{tikzpicture}
			\begin{feynman}
				\vertex (o) ;
				\vertex [above left= 1.1cm of o] (a) {\(f\)};
				\vertex [below left= 1.1cm of o] (b) {\(\bar f\)};
				\vertex [above right= 1.1cm of o] (c) {\(\chi\)};
				\vertex [below right= 1.1cm of o] (d) ;
				\vertex [below right= 1.1cm of d] (e) {\(\chi^\dagger\)};
				\vertex [above right= 1.1cm of d] (g) {\(g\)}; 
				\diagram* {
					(a) -- [fermion,thick] (o),
					(o) -- [fermion,thick] (b),
					(o) -- [scalar,thick] (c),
					(o) -- [scalar,thick] (d),
					(d) -- [scalar,thick] (e),
					(g) -- [graviton,thick] (d)
				};
				~~~~~~\end{feynman}
		\end{tikzpicture}
	\end{minipage}
	\hspace{1cm}
	\begin{minipage}[c]{.24\textwidth}
		\begin{tikzpicture}
			\begin{feynman}
				\vertex (o) ;
				\vertex [above left= 1.1cm of o] (a) {\(f\)};
				\vertex [below left= 1.1cm of o] (b) {\(\bar f\)};
				\vertex [above right= 1.1cm of o] (c) {\(\chi\)};
				\vertex [below right= 1.1cm of o] (d) {\(\chi^\dagger\)};
				\vertex [right= 1.1cm of o] (g) {\(g\)}; 
				\diagram* {
					(a) -- [fermion,thick] (o),
					(o) -- [fermion,thick] (b),
					(o) -- [scalar,thick] (c),
					(o) -- [scalar,thick] (d),
					(g) -- [graviton,thick] (o)
				};
				~~~~~~\end{feynman}
		\end{tikzpicture}
	\end{minipage}
	\caption{Feynman diagrams for graviton bremsstrahlung production from UV freeze-in DM production. }
	\label{fig:UV_freeze_in_diagram}
	\end{figure}
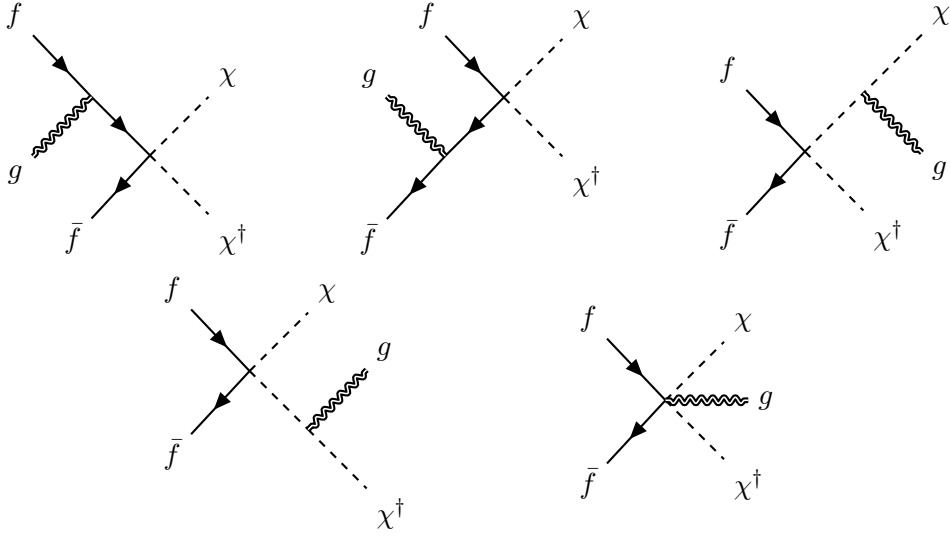

In this section, we calculate the GW spectrum arising from graviton bremsstrahlung during DM production via the UV freeze-in mechanism. The relevant Feynman diagrams are shown in Figure \ref{fig:UV_freeze_in_diagram}.
The squared amplitude for this process is given by:
\begin{equation}
	|\mathcal{M}|^2 = \frac{s}{4 M_P^2 \Lambda^2}
\end{equation}
where \(s\) is the squared center-of-mass energy, \(\Lambda\) is the cutoff scale of the effective theory, and \(M_P\) is the reduced Planck mass. Considering that UV freeze-in typically occurs near the reheating temperature, we work in the relativistic limit and drop all masses. We show the details of the calculation in the appendix \ref{sec:squared amplitude}. 

The evolution of the GW energy density is governed by the Boltzmann equation:
\begin{equation}
	\frac{d \rho_{GW}}{d t} + 4 H \rho_{GW} = C_{f \bar{f} \to \chi \chi^\dagger g}
\end{equation}
where \( C_{f \bar{f} \to \chi \chi^\dagger g} \) is given by:
\begin{equation}
	C_{f \bar{f} \to \chi \chi^\dagger g} = g_f g_{\bar f} \int d \Pi_f d\Pi_{\bar{f}} d\Pi_\chi d\Pi_{\chi^\dagger} d\Pi_g (2\pi)^4 \delta^4(p_f+ p_{\bar{f}} - p_\chi - p_{\chi^\dagger} - p_g) |\mathcal{M}|^2 {\cal F}_f {\cal F}_{\bar{f}} \omega
\end{equation}
where \(\omega \) is the energy of the graviton, \( d\Pi_i \) denotes the invariant phase space element for the particle \( i \), \( g_f= g_{\bar f} = 2 \) are the degrees of freedom for the fermions, and \( {\cal F}_f, {\cal F}_{\bar{f}} \) are the distribution functions of the fermions, which are assumed to be the Boltzmann distributions. Integrating over all the phase space except for the energy of the graviton, and differentiating with respect to the energy of the graviton, we get 
\begin{equation}
	\begin{aligned}
		\frac{d}{d \omega}C_{f \bar{f} \to \chi \chi^\dagger g} = \frac{1}{(2\pi)^7}\frac{T^6 \omega^2}{M_P^2 \Lambda^2} G_{1,3}^{3,0}\left(\frac{\omega^2}{T^2}\Bigg|
\begin{array}{c}
 1 \\
 0,2,3 \\
\end{array}
\right) \\
	\end{aligned}
\end{equation}
where \( G_{1,3}^{3,0} \) is the Meijer G-function.  A detailed derivation is provided in the appendix \ref{sec:collision_terms}. The yield of the gravitons differentiating with respect to the energy of the graviton is given by:
\begin{equation}
	\frac{d}{d\omega_E} Y_{GW} ^\infty = \int _{T_{E}} ^{T_R} dT \left(\frac{45}{2\pi ^2 g_* ^S(T)  }\right)^{4/3} \frac{\sqrt{90}}{\pi \sqrt{g_* ^\rho(T) }}\frac{1}{(2\pi)^7}\frac{ \omega_E ^2 T}{M_P \Lambda^2 T_E^2} G_{1,3}^{3,0}\left(\frac{\omega_E^2}{T_E^2}\Bigg|
\begin{array}{c}
 1 \\
 0,2,3 \\
\end{array}
\right) \\
\end{equation}

The production of gravitons begins at the reheating temperature \(T_R\) and ends at the temperature \(T_E\). The yield of DM produced through the UV freeze-in mechanism satisfies \(Y_{DM}(T = 0.01 T_R) = 0.99 Y_{DM}^{\infty}\), which means that DM production is almost finished when the temperature drops to \(0.01 T_R\). Therefore we take \(T_E = 0.01 T_R\).  $\omega_E$ is the energy of the graviton at the temperature \(T_E\), we have taken into account the energy redshift of the gravitons in the above equation. During the UV freeze-in process \( g_*^S(T)= g_* ^\rho (T) = 106.75\),  assuming that the particle \(f \) is a SM particle. Integrating the above equation with respect to the temperature from \(T_E=0.01T_R\) to \(T_R\), we obtain the yield of the gravitational waves per energy interval at the temperature of \(T_E\):
\begin{equation}
	\frac{d Y_{GW}^\infty}{d \omega_E} = \left(\frac{45}{2\pi ^2 g_* ^S(T_E)  }\right)^{4/3} \frac{\sqrt{90}}{ \sqrt{g_* ^\rho(T_E) }}\frac{1}{(2\pi)^8}\frac{ \omega_E ^2 T_R ^2}{M_P \Lambda^2 T_E^2} G_{1,3}^{3,0}\left(\frac{\omega_E^2}{T_E^2}\Bigg|
\begin{array}{c}
 1 \\
 0,2,3 \\
\end{array}
\right) \\
\end{equation}

\begin{figure}[t]
\centering
\includegraphics[width=1.0\textwidth]{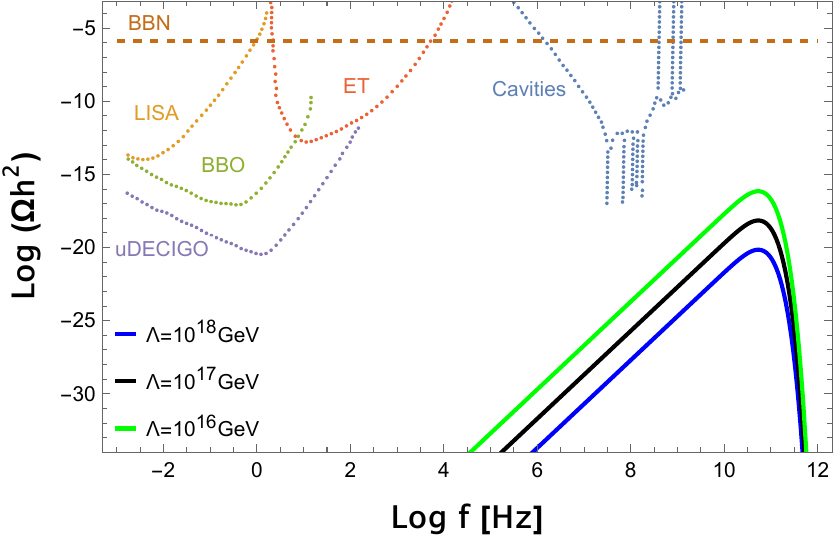}
\caption{The present-day GW spectrum arising from grivoton bremsstrahlung during the DM UV freeze-in. We show the GW spectrum for different values of the cutoff scale \(\Lambda\) with \(T_R = 10^{16}\,\mathrm{GeV}\). We take \(\Lambda = 10^{16}\,\mathrm{GeV}\) (green), \(10^{17}\,\mathrm{GeV}\) (black), and \(10^{18}\,\mathrm{GeV}\) (blue). The corresponding DM mass is \(m_\chi = 1.5 \mathrm{keV} \) (green), \(150 \mathrm{keV}\) (black), and \(15 \mathrm{MeV}\) (blue) respectively.}
\label{fig:GW_spectrum_UV}
\end{figure}

Using the relation \( \omega_E = 2\pi f \frac{a(T_0)}{a(T_E)} \) , one obtains the present-day gravitational wave spectrum as
\begin{equation}
	\begin{aligned}
	\Omega_{GW}(f) = & \frac{1}{\rho_{cr,0}} \frac{d Y_{GW}^\infty}{d\ln \omega_E} S^{4/3}(T_0) 
	= \frac{T_0^4}{\rho_{cr,0}}  \left(\frac{g_*^S (T_0)}{g_*^S (T_E)} \right)^{1/3} \times \\ & \frac{\sqrt{90}}{ \sqrt{g_*^\rho(T_E)}}\frac{1}{(2\pi)^5}\frac{T_R ^2 T_E}{M_P \Lambda^2} \frac{f^3}{T_0^3}  G_{1,3}^{3,0}\left(\frac{(2\pi f)^2}{T_0^2} \left(\frac{g_*^S (T_E)}{g_*^S (T_0)}\right)^{\frac{2}{3}}\Bigg|
\begin{array}{c}
 1 \\
 0,2,3 \\
\end{array}
\right) \\
\end{aligned}
\end{equation}
To reproduce the observed DM relic density, \(\Omega_{DM} h^2 = 0.12\), the reheating temperature \(T_R\), the cutoff scale \(\Lambda\), and the mass of the particle \(\chi\) must satisfy the following relation: \(m_\chi T_R / \Lambda^2 \approx 1.5 \times 10^{-22}\). We fix the reheating temperature to be \(10^{16}\,\mathrm{GeV}\) and vary the cutoff scale \(\Lambda\) to obtain the GW spectrum, which is shown in the Fig.~\ref{fig:GW_spectrum_UV}.
The peak frequency located at \(f_{\text{peak}} \simeq 5.35 \times 10^{10}\,\mathrm{Hz}\) is independent of the cutoff scale \(\Lambda\) and the reheating temperature \(T_R\). The maximum value of the spectrum \(\Omega_{GW}(f_{peak}) \simeq 7.1 \times 10^{-17} \), \(7.1 \times 10^{-29} \), and \(7.1 \times 10^{-21} \) for \(\Lambda = 10^{16}\,\mathrm{GeV}\), \(10^{17}\,\mathrm{GeV}\), and \(10^{18}\,\mathrm{GeV}\) respectively. We get the GW relic density at the peak frequency as:
\begin{equation}
	\Omega_{GW} h^2 (f_{peak}) = 7.1 \times 10^{-17} \left( \frac{T_R}{10^{16}\,\mathrm{GeV}} \right)^3 \left( \frac{\Lambda^2}{10^{32}\,\mathrm{GeV^2}} \right)^{-1}
\end{equation}
which means that the reheating temperature can effectively affect the GW spectrum at the peak frequency.

\section{Conclusion}

In this study, we present a systematic investigation of DM production via the freeze-in mechanism, exploring both the infrared (IR) and ultraviolet (UV) scenarios and their corresponding GW signatures. 
Through detailed analytical derivations, we establish the evolution equations governing the graviton yield and assess their sensitivity to critical model parameters, including coupling constants, mediator and DM particle masses, and the reheating temperature of the early universe. Our findings demonstrate that the GW spectra arising from graviton bremsstrahlung exhibit unique features, which serve as discriminative signatures of the specific freeze-in scenario—whether IR- or UV-dominated. These results underscore the potential of high-frequency gravitational wave observations as a probe of non-thermal DM production mechanisms.
Although the sensitivities of current and proposed GW observatories, including LISA, Einstein Telescope (ET), Big Bang Observer (BBO), ultimate DECIGO (uDECIGO), resonant cavities, and cosmological constraints from Big Bang Nucleosynthesis (BBN), are  insufficient to detect these signals, our work clearly delineates the theoretical benchmarks and sensitivities necessary for future detectors. GW astronomy remains a promising, though presently challenging, avenue for testing theoretical frameworks and parameter spaces associated with freeze-in dark matter. Advancements in GW detection technologies will be essential for probing these subtle yet profound signatures of physics beyond the Standard Model.
			
\begin{acknowledgments}
This work was supported in part by the National Key R\&D Program of China under Grant No. 2023YFA1607104, by the National Natural Science Foundation of China under Grants No. 11775025 and No. 12175027, and by the Fundamental Research Funds for the Central Universities under Grant No. 2017NT17.
\end{acknowledgments}

\appendix 
\appendix
\section{Calculation of the squared amplitude}\label{sec:squared amplitude}
In this appendix, we provide the details of the calculation of the squared amplitude for the processes discussed in the main text.
\subsection{Feynman rules for the graviton interaction}
First, we present the relevant Feynman rules, which have been extensively discussed in previous literature concerning graviton interactions \cite{murayama2025observing, barman2023gravitational, datta2024probing, choi1995factorization, Xu:2025wjq, Garcia-Cely:2024ujr,Choi:2024ilx}.

\begin{figure}[htbp]
    \centering
	\begin{minipage}{.24\textwidth}
        \begin{tikzpicture}
            \begin{feynman}
                \vertex (b) at (-2,-1) {\(\phi\)};
                \vertex (c) at (2,-1) {\(\phi\)};
                \vertex (d) at (0,2) {\(h_{\mu\nu}\)};
                \vertex [dot] (v) at (0,0) {};
                \diagram* {
                    (b) -- [scalar,thick,momentum=$p_1$] (v) -- [scalar,thick,momentum=$p_2$] (c),
                    (v) -- [graviton,thick] (d)
                };
            \end{feynman}
        \end{tikzpicture}
       \vspace{3pt}
    \footnotesize $- i\frac{\kappa}{2}\,[p_{1\mu}p_{2\nu}+p_{1\nu}p_{2\mu}]$
	\end{minipage}
	\hfill
	\begin{minipage}{.24\textwidth}
        \begin{tikzpicture}
            \begin{feynman}
                \vertex (b) at (-2,-1) {\(\chi\)};
                \vertex (c) at (2,-1) {\(\chi\)};
                \vertex (d) at (0,2) {\(h_{\mu\nu}\)};
                \vertex [dot] (v) at (0,0) {};
                \diagram* {
                    (b) -- [scalar,thick,momentum=$p_1$] (v) -- [scalar,thick,momentum=$p_2$] (c),
                    (v) -- [graviton,thick] (d)
                };
            \end{feynman}
        \end{tikzpicture}
	 \vspace{3pt}
    \footnotesize $- i\frac{\kappa}{2}\,[p_{1\mu}p_{2\nu}+p_{1\nu}p_{2\mu}]$
	\end{minipage}
	\hfill
	\begin{minipage}{.24\textwidth}
        \begin{tikzpicture}
            \begin{feynman}
               \vertex (b) at (-2,-1) {\(\psi\)};
                \vertex (c) at (2,-1) {\(\psi\)};
                \vertex (d) at (0,2) {\(h_{\mu\nu}\)};
                \vertex [dot] (v) at (0,0) {};
                \diagram* {
                    (b) -- [scalar,thick,momentum=$p_1$] (v) -- [scalar,thick,momentum=$p_2$] (c),
                    (v) -- [graviton,thick] (d)
                };
            \end{feynman}
        \end{tikzpicture}

	 \vspace{3pt}
    \footnotesize $-\frac{\kappa}{2}\frac{i}{4}\big[(p_1+p_2)_\mu\gamma_\nu+(p_1+p_2)_\nu\gamma_\mu\big]$
	\end{minipage}
	\caption{Feynman rules for the graviton interaction with real scalar, complex scalar and fermion fields. }
    \label{fig:graviton_vertex}
\end{figure}
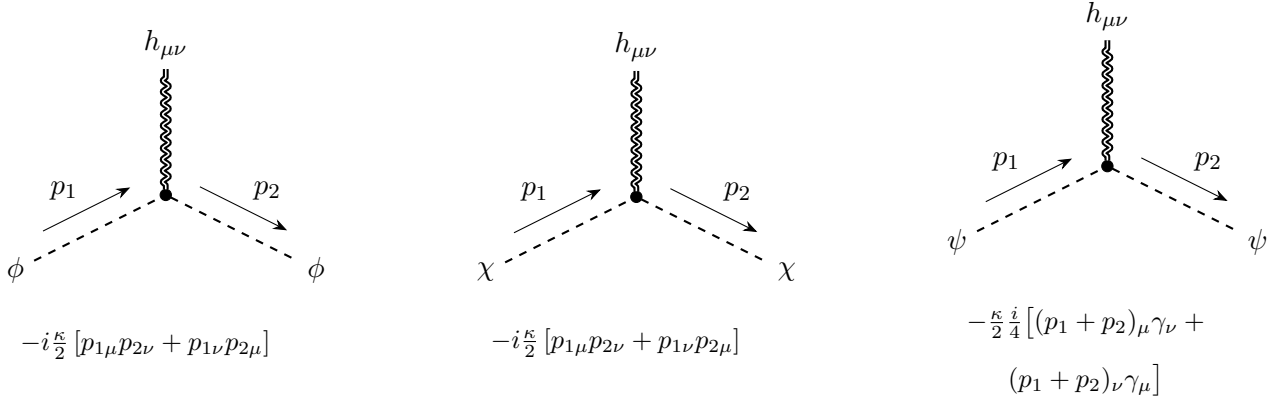

The graviton interacts with other particles through the following effective interaction Lagrangian \cite{Nakayama:2018ptw}:
\begin{equation}
	\mathcal{L}_{int} = -\frac{\kappa}{2} h_{\mu \nu} T^{\mu \nu}
\end{equation}
where \(h_{\mu \nu}\) is the graviton field, \(\kappa = \sqrt{2}/M_P \), and \(T^{\mu \nu}\) is the energy-momentum tensor of the matter fields. For the scalar field \(\phi\), comlex scalar field \(\chi\), and fermion field \(\psi\), their energy-momentum tensors are given by \cite{deAquino:2011ix}:
\begin{equation}
	\begin{aligned}
		T_{\mu \nu}^\phi &= \partial_\mu \phi \partial_\nu \phi - \eta_{\mu \nu} \mathcal{L}_\phi, \\
		T_{\mu \nu}^\chi &= \partial_\mu \chi^\dagger \partial_\nu \chi + \partial_\mu \chi \partial_\nu \chi^\dagger - \eta_{\mu \nu} \mathcal{L}_\chi, \\
		T_{\mu \nu}^\psi &= \frac{i}{4} \bar{\psi} \left( \gamma_\mu \overleftrightarrow{\partial}_\nu + \gamma_\nu \overleftrightarrow{\partial}_\mu \right) \psi - \eta_{\mu \nu} \mathcal{L}_\psi
	\end{aligned}
\end{equation}
where \( \mathcal{L}\) are the Lagrangians for the corresponding fields, and \(\bar{\psi}\overleftrightarrow{\partial}_\mu \psi \equiv \bar{\psi} \partial _\mu \psi - (\partial _\mu \bar{\psi }) \psi\). The polarization tensor of the graviton \(\varepsilon^{\mu\nu}\) satisfies the following conditions \cite{gross1968low, gleisberg2003helicity}:
\begin{equation}
    \begin{aligned}
        \varepsilon ^{i \ \mu \nu} &= \varepsilon ^{i\ \nu \mu} \;\;\; \text{symmetric}\\
        \omega_\mu \varepsilon ^{i \ \mu \nu} &= 0 \;\;\;\; \; \;\;\;\text{transverse}\\
        \eta _{\mu \nu} \varepsilon ^{i \ \mu \nu} &= 0 \;\;\;\; \; \;\;\;\text{traceless}\\
        \varepsilon ^{i \ \mu \nu}\varepsilon ^{j\ *}_{\ \mu \nu} &= \delta ^{ij} \;\;\;\;\;\; \text{orthonormal}
    \end{aligned}
\end{equation}
where \(\omega_\mu\) is the four-momentum of the graviton. Due to the traceless condition, the non-zero interaction vertices are shown in Figure \ref{fig:graviton_vertex}. 
The polarization sum of the graviton is given by:
	\begin{equation}
    \sum_i \varepsilon _{i}^{\alpha \beta}\varepsilon _{i}^{* \mu \nu} = \frac{1}{2} \left( \hat{\eta}^{\alpha \mu}\hat{\eta}^{\beta \nu}+\hat{\eta}^{\alpha \nu}\hat{\eta}^{\beta \mu}-\hat{\eta}^{\mu \nu} \hat{\eta}^{\alpha \beta} \right)
\end{equation}
with 
\begin{equation}
    \hat{\eta}^{\mu \nu } \equiv \eta^{\mu \nu} - \frac{\omega^\mu \bar \omega ^\nu+\omega^\nu \bar \omega ^\mu}{\omega \cdot \bar \omega} 
\end{equation}
where \(\bar \omega ^\mu \equiv (\omega , - \vec{\omega})\). We use \( \omega \) to denote the energy of the graviton, \(\vec{\omega}\) to denote its spatial momentum, \(\omega^\mu\) to denote its four-momentum and \(\omega\cdot \bar{\omega}\) is the scalar product of two four-vectors.

\subsection{Squared amplitude for the IR freeze-in graviton bremsstrahlung process}

The amplitude of the fourth diagram in Figure \ref{fig:IR_freeze_in_diagram} is zero for the traceless condition of the graviton. The amplitude of the first three diagrams in Figure \ref{fig:IR_freeze_in_diagram} is given by:
\begin{equation}
    \begin{aligned}
        i\mathcal{M}_1 
        &= -i\frac{\kappa}{2}\beta \bar u(p_1) p_1^\mu \gamma^\nu \frac{p\cdot\gamma +2m_\psi}{2p_1\cdot \omega} v(p_2)  \varepsilon_{\mu \nu} ^*(\omega)\\
        i\mathcal{M}_2
        &=-i\frac{\kappa}{2}\beta \bar u(p_1) \frac{p\cdot\gamma }{2p_2\cdot \omega}  p_2^\mu \gamma^\nu v(p_2)  \varepsilon_{\mu \nu} ^*(\omega)\\
		i\mathcal{M}_3
		&=i \frac{\kappa}{2}\beta \bar u(p_1) v(p_2) \frac{p^\mu p^\nu}{p\cdot \omega} \varepsilon^*_{\mu \nu}(\omega)
    \end{aligned}
\end{equation}
where \(p\), \(p_1\), and \(p_2\) are the four-momenta of the particles \(\phi\), \(\psi\), and \(\bar{\psi}\) respectively. In the center-of-mass frame of the particle \(\phi\), the four-momenta scalar products are given by:
\begin{equation}
	\begin{aligned}
& p \cdot p = m_p^2 \\
& p \cdot p_1 = m_p E_\psi \\
& p \cdot p_2 = m_p (m_p - E_\psi - \omega) \\
& p \cdot \omega = m_p \omega \\
& p_1 \cdot \omega = m_p \left( \omega + E_\psi \right) - m_p^2/2 \\ 
& p_1 \cdot p_2 = \frac{m_p}{2}(m_p - 2\omega) - m_\psi ^2\\
& p_2 \cdot \omega = \frac{m_p}{2} (m_p - 2 E_\psi)
\end{aligned}
\end{equation}
where \(E_\psi\) is the energy of the particle \(\psi\). Using the polarization sum of the graviton and the above four momenta scalar products, we obtain the total squared amplitude in the limit of \(m_\psi \to 0\):
\begin{equation}
    |\mathcal{M}|^2  = \frac{\beta^{2} m_{\phi}^{2} \kappa ^2}{4}\left(1-\frac{2 \omega}{m_{\phi}}\right)\left[2-\frac{2 m_{\phi}}{\omega}+\left(\frac{m_{\phi}}{\omega}\right)^{2}\right]
\end{equation}
The calculation was performed with the help of the FeynCalc \cite{mertig1991feyn, shtabovenko2016new, shtabovenko2020feyncalc, shtabovenko2025feyncalc}.

\subsection{Squared amplitude for the UV freeze-in graviton bremsstrahlung process}
The amplitude of the last diagram in Figure \ref{fig:UV_freeze_in_diagram} is zero for the traceless condition of the graviton. The amplitude of the first four diagrams in Figure \ref{fig:UV_freeze_in_diagram} is given by:
\begin{equation}
    \begin{aligned}
i\mathcal{M}_1 
        &= \frac{\kappa}{2}\frac{i}{\Lambda} \bar u (p_2) \frac{(p_3+p_4)\cdot \gamma}{2p_1 \cdot \omega} p_1^\mu \gamma^\nu u(p_1) \varepsilon_{\mu \nu} ^*\\
i\mathcal{M}_2 
        &=\frac{\kappa}{2}\frac{i}{\Lambda}\bar u (p_2) p_2 ^\mu \gamma^\nu \frac{(p_3+p_4)\cdot \gamma }{2 p_2 \cdot \omega}u(p_1) \varepsilon_{\mu \nu} ^*\\
i\mathcal{M}_3
       &=\frac{\kappa}{2}\frac{-i}{\Lambda} \bar u(p_2)u(p_1)\frac{2 p_3 ^\mu p_3 ^\nu}{2 p_3 \cdot \omega} \varepsilon_{\mu \nu}^*   \\
i\mathcal{M}_4 
        &= \frac{\kappa}{2}\frac{-i}{\Lambda} \bar u(p_2) u(p_1) \frac{2 p_4 ^\mu p_4 ^\nu}{2 p_4 \cdot \omega} \varepsilon_{\mu \nu}^*
    \end{aligned}
\end{equation}
where \(p_1\), \(p_2\), \(p_3\), and \(p_4\) are the four-momenta of the particles \(f\), \(\bar{f}\), \(\chi\), and \(\chi^\dagger\) respectively. In the center-of-mass frame of \(f\) and \(\bar{f}\), the four-momenta of each particle are given by:
\begin{equation}
	\begin{aligned}
		p_1^{\mu} &= (\sqrt{s}/2, \vec{p}) \\
		p_2^{\mu} &= (\sqrt{s}/2, -\vec{p}) \\
		p_3^{\mu} &= (E_3, \vec{p}_3) \\
		\omega^{\mu} &= (\omega, \vec{\omega}) \\
		p_4^{\mu} &= (\frac{s}{2}-E_3-\omega, -\vec{p_3}-\vec{\omega}) \\
    \end{aligned}
\end{equation}
Among the four-momenta, only \(p_1^\mu\), \(p_3^\mu\), and \(\omega^\mu\) are independent. The scalar products between them are given by:
\begin{equation}
	\begin{aligned}
& p_1 \cdot \omega = \omega \sqrt{s}/2 - |\vec{p}| \omega \cos\theta \\
& p_3 \cdot \omega = E_3 \omega - |\vec{p}_3| \omega \cos\varphi \\
& p_1 \cdot p_3 = E_3 \sqrt{s}/2 - |\vec{p}_1| |\vec{p}_3| \cos(\varphi-\theta)\\
	\end{aligned}
\end{equation}
where \(s\) is the squared energy in the center-mass-frame, \(\theta\) is the angle between \(\vec{p_1}\) and \(\vec{\omega}\), \(\varphi\) is the angle between \(\vec{p}_3\) and \(\omega\), and \( \cos \varphi = \left(s^2 - 2 \sqrt{s}(E_3+\omega) + 2E_3 \omega \right)/(2|\vec{p}_3| \omega) \). The rest of the scalar products can be obtained using the conservation of 4-momentum. Then we obtain the total squared amplitude in the massless limit
\begin{equation}
	|\mathcal{M}| = \frac{\kappa^2}{4}\frac{s}{2\Lambda ^2}
\end{equation} 
This result is obtained with the help of FeynCalc, and we have averaged over the spins of the initial particles.

\section{Calculation of the Collision terms} \label{sec:collision_terms}
In this section, we provide the details of the calculation of the collision terms for the processes discussed in the main text. 

\subsection{Collision term for the IR freeze-in graviton bremsstrahlung process}
The collision term for the process \(\phi \to \psi \bar{\psi} g\) is given by:
\begin{equation}
	C_{\phi \to \psi \bar{\psi} g} = g_\phi \int d\Pi_\phi d\Pi_\psi d\Pi_{\bar{\psi}} d\Pi_g (2\pi)^4 \delta^4(p_\phi - p_\psi - p_{\bar{\psi}} - p_g) |\mathcal{M}|^2 f_\phi \omega
\end{equation}
First, we integrate the Lorentz-invariant phase space for the three-body final state:
\begin{equation}
	\begin{aligned}
\int d \Pi_{\text{LIPS}} ^{(3)} &= \int \frac{d^3 p_\psi}{(2\pi)^3 2 E_\psi} \frac{d^3 p_{\bar{\psi}}}{(2\pi)^3 2 E_{\bar{\psi}}} \frac{d^3 p_g}{(2\pi)^3 2 \omega} (2\pi)^4 \delta^4(p_\phi - p_\psi - p_{\bar{\psi}} - p_g) \\
&= \int \frac{d \Omega_\psi p_\psi^2 d p_\psi}{(2\pi)^3 2 E_\psi} \frac{d\Omega_g \omega^2 d\omega}{(2\pi)^3 2 \omega} \frac{1}{2E_{\bar{\psi}}} 2\pi \delta(m_\phi - E_\psi - E_{\bar{\psi}} - \omega) \\
	\end{aligned}
\end{equation}	
where \(d\Omega_\psi\) and \(d\Omega_g\) are the solid angles of the particles \(\psi\) and the graviton, respectively. Here, \(p_\psi\) and \(\omega\) may refer either to the magnitudes of their three-momenta or to their four-momenta, depending on the context. In particular, the energy of the graviton is equal to the magnitude of its three-momentum. We hope this will not cause confusion. And, 
\(E_{\bar \psi} = \sqrt{(\vec{p}_\psi+ \vec{\omega})^2+m_\psi ^2 } \), where we have chosen the center-of-mass frame of the particle \(\phi\) as the result of the integration is independent of the choice of the frame. If the squared amplitude does not depend on the angle between the initial and final particles momenta, we can write the differential solid angle element as: 
\begin{equation}
	d \Omega_\psi d\Omega_g =8 \pi^2 d\cos \theta 
\end{equation}
where \(\theta\) is the angle between the three-momenta of the particles \(\psi\) and graviton and there is a \(\cos \theta \) in the \(\delta\) function
\begin{equation}
	\delta \left(m_\phi - E_\psi - \sqrt{p_\psi^2 +\omega^2 + 2 p_\psi \omega \cos \theta + m_\psi^2} -\omega \right)
\end{equation} 
so we can integrate over the \(\cos \theta\) using the \(\delta\) function, which gives:
\begin{equation}
	\begin{aligned}
		\int d \Pi_{\text{LIPS}} ^{(3)}=\int \frac{dE_\psi d\omega}{32\pi^3} 
	\end{aligned}
	\label{eq:integrated_LIPS}
\end{equation}	
where we have used \(p_\psi dp_\psi = E_\psi dE_\psi\). Next, we determine the integration ranges for \(\omega\) and \(E_\psi\). The minimum value of \(\omega\) is zero, which occurs when \(\psi\) and \(\bar{\psi}\) move in opposite directions with equal momenta. The maximum value of \(\omega\) is \((m_\phi^2 - 4 m_\psi^2)/(2m_\phi)\), which is reached when \(\psi\) and \(\bar{\psi}\) are at rest with respect to each other. Once the values of \(\omega\) are determined, the value of \(E_\psi\) can be obtained from the energy and momentum conservation conditions:
\begin{equation}
	(m_\phi-E_\psi-E_\omega)^2  = \omega^2+E_\psi^2+2\omega p_\psi \cos \theta
\end{equation}
where \(\theta\) is the angle between the three-momenta of the particles \(\psi\) and graviton. The maximum value of \(E_\psi\) is reached when \( \cos \theta = 1\), which gives:
\begin{equation}
	E_\psi^{\text{max}} = \frac 12 \left( m_\phi -\omega + \omega \sqrt{\frac{m_\phi ^2-2\omega m_\phi-4m_\psi^2}{m_\phi(m_\phi-2\omega)}} \right)
\end{equation}
The minimum value of \(E_\psi\) is reached when \(\cos \theta = -1\), which gives:
\begin{equation}
	E_\psi^{\text{min}} = \frac 12 \left( m_\phi -\omega - \omega \sqrt{\frac{m_\phi ^2-2\omega m_\phi-4m_\psi^2}{m_\phi(m_\phi-2\omega)}} \right)
\end{equation}
In the limit of \(m_\psi \to 0\), the range of \(\omega\) is \(\left[0, \frac{m_\phi}{2}\right]\) and the range of \(E_\psi\) is \(\left[m_\phi/2-\omega, m_\phi/2 \right]\). Since the squared amplitude is independent of the momentum of \(\phi\), we can integrate over the invariant phase space of \(\phi\) together with its distribution function \(f_\phi\), yielding: 
\begin{equation}
	\int \frac{d^3 p_\phi}{(2\pi)^3 2 E_\phi} f_\phi = \frac{1}{(2\pi)^2} \int _{m_\phi} ^{\infty}\sqrt{E_\phi^2 - m_\phi^2} \text{e}^{-E_\phi/T} = \frac{m_\phi T K_1\left(m_\phi /T\right)}{4 \pi ^2}
\end{equation}
where \(K_1\) is the first modified Bessel function of the second kind. Finally, the collision term differentiated with respect to the energy of the graviton \(\omega\) is given by:
\begin{equation}
	\begin{aligned}
		\frac{d}{d\omega}C_{\phi \to \psi \bar{\psi} g} &= g_\phi \frac{m_\phi T K_1\left(m_\phi /T\right)}{4 \pi ^2} \int _{m_\phi/2-\omega}^{m_\phi/2} dE_\psi \frac{1}{32\pi^3} |\mathcal{M}|^2 \omega \\
		&= g_\phi \frac{m_\phi T K_1\left(m_\phi /T\right)}{128 \pi ^5} |\mathcal{M}|^2 \omega^2
	\end{aligned}
	\end{equation}

\subsection{Collision term for the UV freeze-in graviton bremsstrahlung process}
The collision term for the process \(f \bar{f} \to \chi \chi^\dagger g\) is given by:
\begin{equation}
	C_{f \bar{f} \to \chi \chi^\dagger g} = g_f g_{\bar{f}} \int d\Pi_f d\Pi_{\bar{f}} d\Pi_\chi d\Pi_{\chi^\dagger} d\Pi_g (2\pi)^4 \delta^4(p_f + p_{\bar{f}} - p_\chi - p_{\chi^\dagger} - p_g) |\mathcal{M}|^2 {\cal F}_f {\cal F}_{\bar{f}} \omega
\end{equation}
The integration over the Lorentz-invariant phase space for the three-body final state is same as Eq. \eqref{eq:integrated_LIPS}, but the integration ranges for \(\omega\) and \(E_f\) are \(\left[0, \sqrt{s}/2\right]\) and \(\left[\sqrt{s}/2-\omega, \sqrt{s}/2 \right]\) in the massless limit, respectively, where \(s = (p_f + p_{\bar{f}})^2\) is the center-of-mass energy squared. In the relativistic limit, the invariant phase space for the initial particles is given by:
\begin{equation}
	 \frac{d^3 p_f}{(2\pi)^3 2 E_f} \frac{d^3 p_{\bar{f}}}{(2\pi)^3 2 E_{\bar{f}}} = \frac{1}{2(2\pi)^4}E_f E_{\bar{f}} d\cos \theta dE_f dE_{\bar{f}}
\end{equation}
where \(\theta\) is the angle between the three-momenta of the particles \(f\) and \(\bar{f}\). It is convenient to make the following change of variables \cite{edsjo1997neutralino, elahi2015ultraviolet} 
\begin{equation}
	E_+ \equiv E_f + E_{\bar{f}}, \quad E_- \equiv E_f - E_{\bar{f}}, \quad s= 2E_f E_{\bar{f}}(1-\cos \theta) 
\end{equation}
Accordingly, the phase space volume element can be expressed in terms of these new variables as follows:
\begin{equation}
	\frac{1}{2(2\pi)^4}E_f E_{\bar{f}} d\cos \theta dE_f dE_{\bar{f}} = \frac{1}{8(2\pi)^4} dE_- dE_+ ds
\end{equation}
where \( |E_-| \leq \sqrt{E_+ ^2 -s} \), \(E_+ \geq \sqrt{s}\) and for fixed \(\omega\), \(\sqrt{s} \geq 2 \omega \). Therefore, the integration over the invariant phase space of the initial particles together with their distribution functions \({\cal F}_f\) and \({\cal F}_{\bar{f}}\) yields:
\begin{equation}
	\begin{aligned}
\int d\Pi_{f} d\Pi_{\bar{f}} {\cal F}_f {\cal F} _{\bar{f}} &= \frac{1}{8(2\pi)^4} \int _{(2\omega)^2} ^{\infty} ds \int _{\sqrt{s}} ^{\infty} dE_+ \int _{-\sqrt{E_+^2 -s}}^{\sqrt{E_+^2 -s}} dE_- \text{e}^{-E_+/T} \\
&= \frac{T}{4(2\pi)^4}\int _{(2\omega)^2} ^{\infty} ds \sqrt{s} K_1\left(\sqrt{s}/T\right)
	\end{aligned}
\end{equation}
Finally, the collision term differentiated with respect to the energy of the graviton \(\omega\) is given by:
	\begin{equation}
		\begin{aligned}
		\frac{d}{d\omega}C_{f \bar{f} \to \chi \chi^\dagger g} &= g_f g_{\bar{f}} \frac{T}{4(2\pi)^4} \int _{(2\omega)^2} ^{\infty} ds \sqrt{s} K_1\left(\sqrt{s}/T\right)\frac{1}{32\pi^3} \frac{s}{4M_P ^2 \Lambda^2} \omega^2 \\
		&=\frac{1}{(2\pi)^7}\frac{T^6 \omega^2}{M_P^2 \Lambda^2} G_{1,3}^{3,0}\left(\frac{\omega^2}{T^2}\Bigg|
\begin{array}{c}
 1 \\
 0,2,3 \\
\end{array}
\right) \\
	\end{aligned}
	\end{equation}

\bibliographystyle{apsrev4-1}
\bibliography{GWandDMref}
\end{document}